\documentclass[letterpaper]{article} 
\usepackage{aaai24}  
\usepackage{times}  
\usepackage{helvet}  
\usepackage{courier}  
\usepackage[hyphens]{url}  
\usepackage{graphicx} 
\usepackage{natbib}  
\usepackage{caption} 
\usepackage{booktabs}
\usepackage{multirow}
\usepackage{subcaption}
\usepackage{algorithmicx}
\usepackage[noend]{algpseudocode}
\usepackage[linesnumbered,ruled,vlined]{algorithm2e}
\usepackage{hyperref}

\newcommand{\papername}{MedBlindTuner}

\title{{\papername}: Towards Privacy-preserving Fine-tuning on Biomedical Images with Transformers and Fully Homomorphic Encryption \footnote{Accepted for the presentation at W3PHIAI @The 38th Annual AAAI Conference on Artificial Intelligence 2024} }
\author {
    Prajwal Panzade \textsuperscript{\rm 1},
    Daniel Takabi \textsuperscript{\rm 2},
    Zhipeng Cai \textsuperscript{\rm 1}
}
\affiliations {
    \textsuperscript{\rm 1}Georgia State University \\
    \textsuperscript{\rm 2}Old Dominion University \\
    ppanzade1@student.gsu.edu, takabi@odu.edu, zcai@gsu.edu
}

\usepackage{bibentry}

\begin{document}

\maketitle

\begin{abstract}
Advancements in machine learning (ML) have significantly revolutionized medical image analysis, prompting hospitals to rely on external ML services. However, the exchange of sensitive patient data, such as chest X-rays, poses inherent privacy risks when shared with third parties. Addressing this concern, we propose MedBlindTuner, a privacy-preserving framework leveraging fully homomorphic encryption (FHE) and a data-efficient image transformer (DEiT). MedBlindTuner enables the training of ML models exclusively on FHE-encrypted medical images. Our experimental evaluation demonstrates that MedBlindTuner achieves comparable accuracy to models trained on non-encrypted images, offering a secure solution for outsourcing ML computations while preserving patient data privacy. To the best of our knowledge, this is the first work that uses data-efficient image transformers and fully homomorphic encryption in this domain.
\end{abstract}

\section{Introduction}

In recent years, transformers have emerged as the predominant neural architecture for tasks involving sequential modeling, such as language processing, speech comprehension, and computer vision \cite{vaswani2017attention,devlin2018bert,touvron2021training}. Their exceptional performance primarily derives from their reliance on attention mechanisms and extensive pretraining. Recent studies have also highlighted the promising outcomes of vision transformers (ViT) in the domain of biomedical image classification \cite{regmi2023vision}. The advancement of cloud computing has led many Machine Learning as a Service (MLaaS) providers to facilitate fine-tuning with pretrained transformers on their platforms. However, in handling sensitive data, adherence to privacy regulations like GDPR (General Data Protection Regulation) and CCPA (California Consumer Privacy Act) is imperative for these service providers \cite{ribeiro2015mlaas}.

Within a standard MLaaS system, the client retains ownership of the data, while the ML computational processing is handled by the cloud \cite{liu2021machine}. However, when this data encompasses confidential records like healthcare details, significant privacy concerns arise. For instance, envision a scenario where a hospital intends to utilize Company X's skin cancer prediction service. Even if patients provide consent for accessing their medical data, external sharing of such sensitive information for predictive modeling introduces inherent privacy risks. The act of transmitting this data could potentially lead to breaches or unauthorized access, thereby violating patient confidentiality. Moreover, in cases where a patient denies consent, the computations necessary for model development become unfeasible, thereby impeding progress. To tackle these challenges effectively, the implementation of a robust privacy-preserving framework becomes imperative. Such a framework should ensure the protection of data privacy throughout its transmission, processing, and utilization. Additionally, it must instill a sense of trust among patients regarding the safeguarding of their private data by hospitals, thereby fostering reliance on these healthcare institutions. 

Numerous methodologies for preserving privacy in machine learning through secure multiparty computation (SMC) have been developed, such as SecureML \cite{secureml}, SecureNN \cite{securenn}, and DeepSecure \cite{deepsecure}. While these techniques have proven effective, they typically necessitate extensive communication between the client and server \cite{securenn}. For scenarios requiring reduced communication rounds, fully homomorphic encryption (FHE)-based techniques like CryptoNets \cite{gilad2016cryptonets}, CryptoDL \cite{hesamifard2018privacy}, and ML Confidential \cite{mlconfidential} are often favored. Nonetheless, the predominant focus of research in this domain has tended towards private inference rather than training \cite{reagen2021cheetah,gilad2016cryptonets}. Recent research has introduced encrypted neural network methods \cite{nandakumar2019towards} and privacy-preserving transfer learning approaches, as seen in Glyph \cite{lou2020glyph} and HETAL \cite{lee2023hetal}. However, these existing methodologies within the domain of image classification frequently manifest inefficiencies, being either impractically slow for real-world applications or requiring further refinement in computational procedures.

In response to this issue, we present {\papername}, a privacy-preserving training framework designed specifically for machine learning modeling in the realm of medical image classification.

The contributions of this paper are as follows:
\begin{itemize}
    \item We propose {\papername} framework for training an ML model on FHE-encrypted medical images of the patients, where the computations are performed on encrypted data, preserving the privacy of the patients.
    \item {\papername} is a generalized framework that leverages FHE and DEiT for image classification on 2D medical images.
    \item We provide a thorough experimental analysis and benchmarks of {\papername} for multi-class medical image classification on five different datasets from the MedMNIST project \cite{medmnistv2}.
    \item The implementation of {\papername} demonstrates its capacity to train ML models effectively, preserving privacy without substantially compromising accuracy when compared to their non-encrypted equivalents. Moreover, the implementation of {\papername} does not demand extensive expertise in cryptography. {\papername} will be 
available at \url{https://github.com/prajwalpanzade/MedBlindTuner}.

\end{itemize}
\section{Background Knowledge}
\subsection{Fully Homomorphic Encryption}

Fully Homomorphic Encryption (FHE), introduced by Gentry et al. \cite{gentry2009fully}, is an advanced cryptographic technique that enables computations on encrypted data without the need for decryption. Essentially, it allows a third party to perform computations on encrypted data without accessing the data itself or the resulting computations. FHE has far-reaching implications for privacy and security across various applications, particularly in cloud computing and data outsourcing scenarios.

FHE encompasses several variations, including CKKS (Cheon-Kim-Kim-Song) \cite{cheon2017homomorphic}, BGV (Brakerski-Gentry-Vaikuntanathan) \cite{brakerski2014leveled}, and the BFV (Brakerski-Fan-Vercauteren scheme). Among these, CKKS is gaining popularity due to its ability to handle real numbers. Similar to public key encryption (PKE), the CKKS scheme involves encryption, decryption, and key generation algorithms. However, unlike PKE, CKKS integrates homomorphic addition and multiplication functionalities, allowing operations on ciphertexts.

A concise overview of these algorithms includes the following:

\begin{itemize}
    \item \texttt{KeyGen(1$^\lambda$)}: Generates a public key (\texttt{pk}), a secret key (\texttt{sk}), and an evaluation key (\texttt{evk}).
    \item \texttt{Enc\_{pk}(m)}: Encrypts a message (\texttt{m} $\in$ \texttt{R}) using the public key (\texttt{pk}), resulting in ciphertext \texttt{c}, where \texttt{R} represents a set of real numbers.
    \item \texttt{Dec\_{sk}(c)}: Utilizing the secret key (\texttt{sk}), this operation retrieves the original message \texttt{m} from a given ciphertext \texttt{c}.
    \item \texttt{Add(c$_1$, c$_2$)}: Produces element-wise addition \texttt{Enc(m$_1$+m$_2$)} when provided with ciphertexts \texttt{c$_1$} and \texttt{c$_2$}.
    \item \texttt{Mult\_{evk}(c$_1$, c$_2$)}: Generates element-wise multiplication \texttt{Enc(m$_1$*m$_2$)} for a pair of ciphertexts (\texttt{c$_1$, c$_2$}) and an \texttt{evk}. Both addition and multiplication operations produce ciphertexts, requiring the secret key (\texttt{sk}) for decryption. Machine learning computations rely on multivariate polynomials, and the CKKS scheme supports bootstrapping, enabling the computation of multivariate polynomials of arbitrary degrees \cite{cheon2018bootstrapping}.
\end{itemize}

For further details on the CKKS scheme, comprehensive insights and in-depth discussions can be found in \cite{cheon2017homomorphic} and \cite{cheon2018bootstrapping}.

\subsection{Data-Efficient Image Transformers}

Vision transformer (ViT) has emerged as a promising architecture for image classification tasks \cite{dosovitskiy2020image}. However, historically, achieving competitive performance with ViT models required extensive pretraining on large datasets, setting them apart from convolutional neural networks (CNNs) \cite{lecun1998gradient}. Touvron et al. introduced DeiTs \cite{touvron2021training}, showcasing that these transformers can either match or exceed the performance of state-of-the-art CNNs when exclusively trained on ImageNet. They implemented several modifications to the training methodology, integrating extensive data augmentation techniques such as RandAugment, CutMix, and repeated augmentation. Furthermore, they introduced an innovative distillation procedure that employs a distillation token engaging with other embeddings through self-attention. DEiT marks a significant advancement, establishing transformers as a viable alternative to CNNs in computer vision tasks. DEiT stands out for its capacity to train high-performing transformer models without relying extensively on large datasets.

\subsection{Transfer Learning}
Transfer Learning (TL) is a machine learning technique that utilizes previously acquired knowledge to address related yet distinct problems \cite{pan2009survey}. TL involves retraining a model initially trained on a comprehensive dataset using a smaller secondary dataset \cite{weiss2016survey}. The rationale behind TL lies in the recognition that lower levels of neural networks can identify fundamental and transferable features, such as edges, relevant across various tasks. Consequently, pretrained models serve as valuable starting points, demanding less data to learn task-specific features. In computer vision, TL extensively utilizes large pretrained models like VGG \cite{simonyan2014very}, ResNet \cite{he2016deep}, and EfficientNet \cite{tan2019efficientnet}, pretrained on ImageNet \cite{deng2009imagenet}, subsequently fine-tuned for specialized domains like medical or aerial imaging. By leveraging pretrained features, TL achieves high accuracy even with moderately-sized datasets. The process of fine-tuning a pretrained model proves notably faster and more data-efficient compared to training a model from scratch. TL remains a predominant methodology responsible for breakthrough advancements, particularly in applications constrained by limited training data \cite{weiss2016survey}.

\section{The Proposed Methodhology}
\subsection{Threat Model}
\begin{figure*}[!ht]
    \centering
    \includegraphics[scale=0.5]{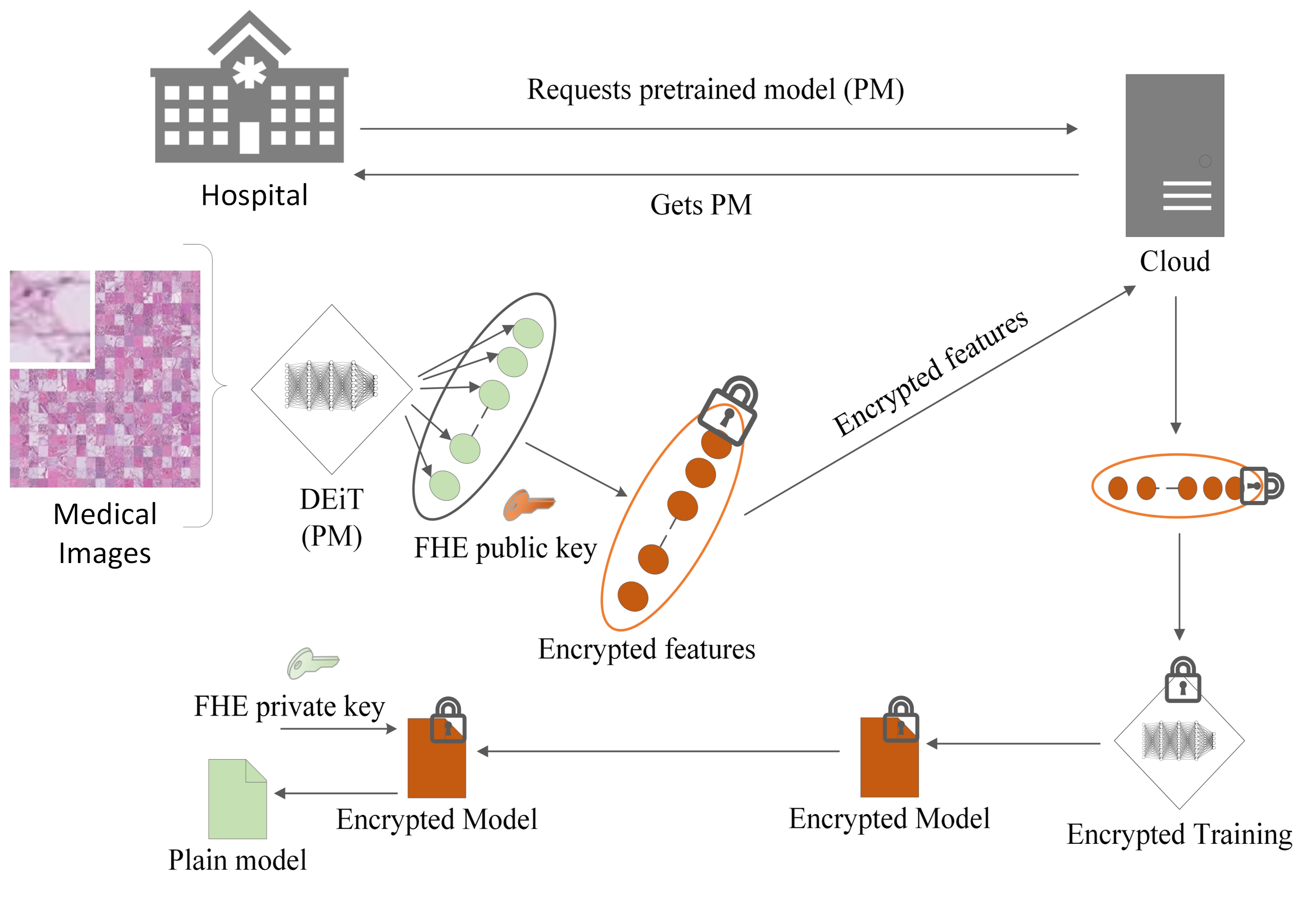}
    \caption{Overview of end-to-end \papername}
    \label{fig:framework}
\end{figure*}

As shown in Figure \ref{fig:framework}, {\papername} consists of two parties: a hospital (client) and a medical ML cloud service provider which we refer to as cloud in this paper. We assume a hospital is seeking ML training and inference services from the ML service provider for the classification of biomedical images. Also, we assume that a hospital has consent to use patients' data for medical analysis. Although the hospital has consent from the patient, they must make sure that the medical analysis happens without hampering the patient's privacy.

\subsection{{\papername}}

Subsequent subsections present computations performed by the hospital and the cloud. \\

\textbf{Hospital.}
The hospital, serving as the custodian of data, seeks cloud-based privacy-preserving services for model training to alleviate computational burdens. Initially, mutual agreement between the hospital and the cloud involves employing a pretrained DEiT model (PM) for performing fine-tuning on medical images. The hospital utilizes the DEiT model to extract features from the dataset obtained from the patients and designated for model training. Following feature extraction, the hospital preprocesses and encrypts these features using CKKS-based FHE with its public key, resulting in encrypted features. These encrypted features are subsequently transmitted to the cloud for further processing.\\

\textbf{Cloud.}
Upon receipt of the encrypted features from the hospital, the cloud utilizes these encrypted features to fine-tune the ML model. The fine-tuning process integrates Nesterov's accelerated gradient (NAG) \cite{nesterov1983method} and encrypted matrix multiplication as suggested in \cite{lee2023hetal} to approximate softmax activation. NAG is well-known for facilitating faster convergence in FHE-based ML computations \cite{crockett2020low}. Since all computations take place on encrypted data, the cloud is not exposed to any raw data. After the encrypted fine-tuning, the ML model is set for inference.  During inference, only encrypted features are required, and the cloud sends the output layer results back to the hospital. Subsequently, the hospital decrypts the results using its private key.\\

\textbf{Assumption.}
Here, it is assumed that the feature extraction, encryption, and decryption processes carried out by the hospital are conducted within a specially designed software interface. This interface enables hospital staff operating the system to perform these cryptographic and feature extraction operations without requiring specialized knowledge of cryptography.\\

\textbf{Security.}
{\papername} provides robust security guarantees during fine-tuning. All features obtained from the hospital are processed in encrypted form in the cloud, preventing adversarial data exposure. Additionally, the FHE schemes used offer quantum-hardened security, safeguarding the original plaintext data and computed model outcomes against unauthorized changes even in the event of compromised infrastructure \cite{creeger2022rise}. 

\section{Experimental Results}
\subsection{Datasets}
We use 5 datasets proposed in MedMNIST2D \cite{medmnistv2} for multiclass image classification as follows:
\begin{itemize}
    \item \textbf{DermaMNIST.} It utilizes the HAM10000 dataset \cite{tschandl2018ham10000,DVN/DBW86T_2018,codella2019skin}, containing 10,015 dermatoscopic images of 7 different diseases, designed for a multi-class classification task. Images are resized from 3$\times$600$\times$450 to 3$\times$28$\times$28 and split into a 7:1:2 ratio for training, validation, and test sets.
    
    \item \textbf{BloodMNIST:} It is derived from a dataset \cite{acevedo2020dataset,acevedo2019recognition} of 17,092 images of normal cells from individuals without specific conditions. It is organized into 8 classes and split 7:1:2 for training, validation, and test sets. Images are resized from 3$\times$360$\times$363 to 3$\times$28$\times$28.
    \item \textbf{Organ\{A,C,S\}MNIST:} They are derived from 3D CT (computed tomography) images from the LiTS dataset \cite{bilic2023liver}, resized and processed into 1$\times$28$\times$28 images for multi-class classification of 11 body organs, differing only in views (Axial, Coronal, Sagittal). It utilizes 115 and 16 CT scans for training and validation and 70 CT scans for the test set.
\end{itemize}

\subsection{Environment Configuration}
To facilitate FHE operations, we employ the HEaaN library \cite{cheon2017homomorphic}, chosen specifically for its built-in support for bootstrapping \cite{cheon2018bootstrapping}. Our implementation utilizes the GPU-accelerated variant of the HEaaN library (0.2.0)\footnote[1]{\url{https://hub.docker.com/r/cryptolabinc/heaan}}, obtained directly from its developers. For feature extraction using pretrained transformers on the hospital-side, our setup relies on Python (3.8.5), PyTorch (2.0.1), TorchVision (0.15.2), NumPy (1.22.2), and Transformers (4.33.1) libraries. The numbers in the brackets show the versions of the software packages used in our experiments. All experiments are performed on a workstation equipped with an Intel Xeon Gold 6230 R processor running at a clock speed of 2.10 GHz and 755 GB of accessible RAM. Additionally, the workstation incorporates an NVIDIA Tesla V100 32 GB GPU and operates on the Ubuntu 20.04.4 OS.  


\subsection{Experiments}
\quad
\textbf{Hospital.}
The hospital employs a DEiT pretrained model to extract features from specific medical image datasets, as previously outlined. The DEiT model version utilized is the deit-base-distilled-patch16-224 \footnote[2]{\url{https://huggingface.co/facebook/deit-base-distilled-patch16-224}} by Facebook research, accessible via the Transformers library. Following feature extraction, the hospital performs dataset partitioning into distinct train, validation, and test sets. Before transmission to the cloud, the training and validation sets undergo encoding and encryption using the ML submodule integrated into the HEaaN library. The employed FHE context setting is FGb, configured with a cyclomatic ring dimension of 2$^{16}$, ensuring a 128-bit security level, as delineated in \cite{cheon2021practical}. Keys for each dataset experiment are generated at the experiment's start and maintained throughout, ensuring consistency across experiments. Notably, the feature extraction duration in the plain domain is not accounted for in this process, as it is considered an offline procedure and does not influence the encrypted fine-tuning demonstration.\\\\

\textbf{Cloud.}
Upon receiving the encrypted training and validation sets, the cloud initializes the ML model for encrypted training using the ML submodule integrated into the HEaaN library. Hyperparameter tuning commences with a batch size of 2048, a learning rate of 1, and 10 epochs. After iterative adjustments to the hyperparameters, optimal configurations are identified and detailed in Table \ref{tab:parameters}. Experimental outcomes, presented in Table \ref{tab:results}, illustrate the results obtained across diverse datasets. Enc training time refers to the duration required for training the encrypted model, while Enc accuracy signifies the test accuracy achieved by the encrypted model. Similarly, Unenc accuracy and Unenc time denote the test accuracy and computation time for the unencrypted model, respectively. The same hyperparameters are used across both encrypted and unencrypted domains to facilitate fair comparison. Results provided in Table \ref{tab:results} highlight the performance of {\papername} for both encrypted and unencrypted models, revealing slight variations in the performance of the encrypted models in comparison to their unencrypted counterparts. This demonstrates the efficacy of the approximation arithmetic methods proposed in \cite{cheon2017homomorphic} and \cite{lee2023hetal} for accurate ML model training, despite the computationally intensive nature of cryptographic FHE computations. However, the advantages of preserving user data privacy without exposing information to the cloud outweigh the computational time.

\subsection{Performance of {\papername}}
\begin{figure*}[!htbp] 
    \centering


\begin{subfigure}{0.42\textwidth}
        \includegraphics[width=\linewidth]{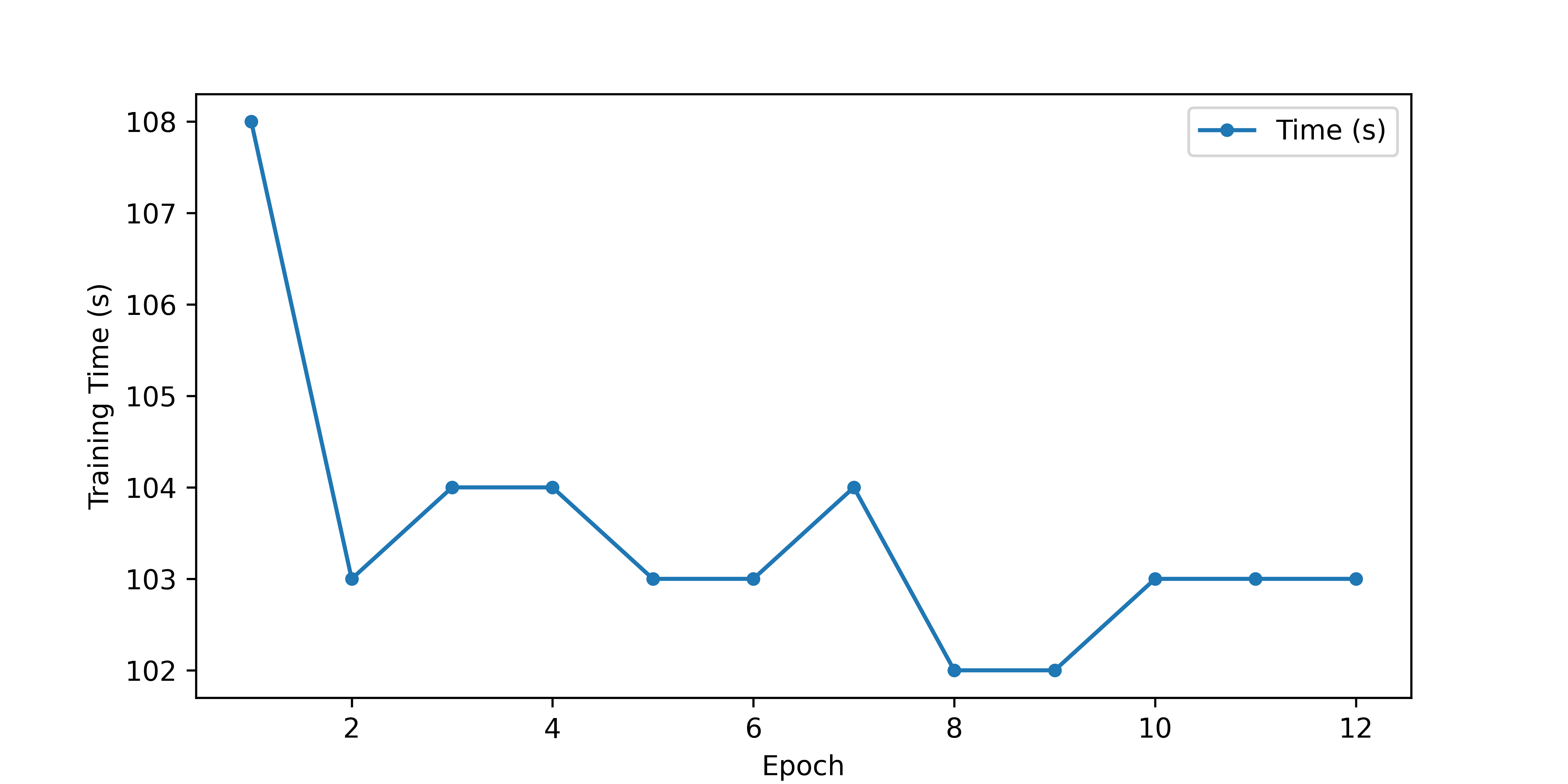}
        \caption{DermaMNIST}
        \label{subfig:sub3}
    \end{subfigure}
    \hfill
    \begin{subfigure}{0.42\textwidth}
        \includegraphics[width=\linewidth]{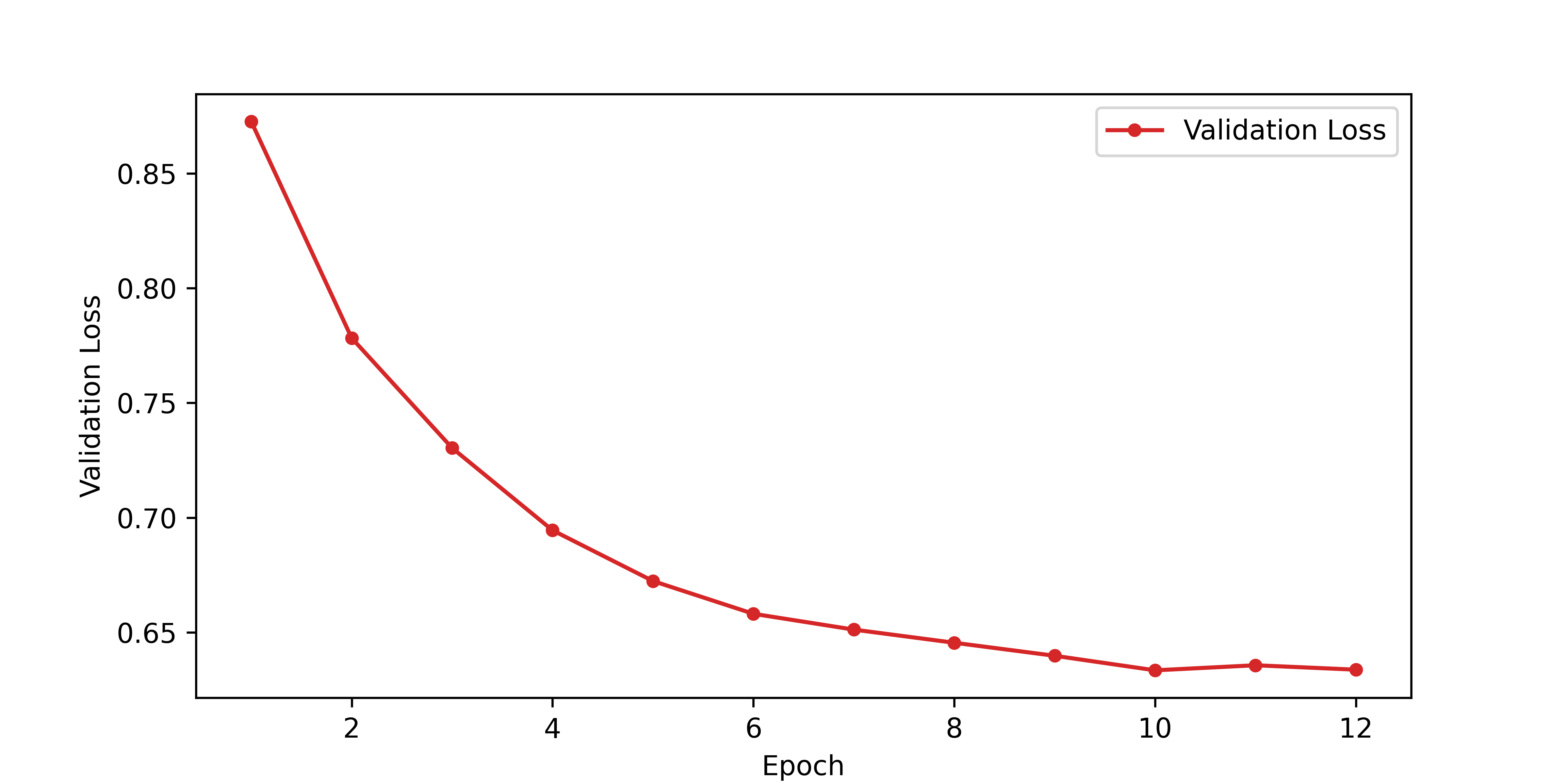}
        \caption{DermaMNIST}
        \label{subfig:sub4}
    \end{subfigure}
\hfill
\begin{subfigure}{0.42\textwidth}
        \includegraphics[width=\linewidth]{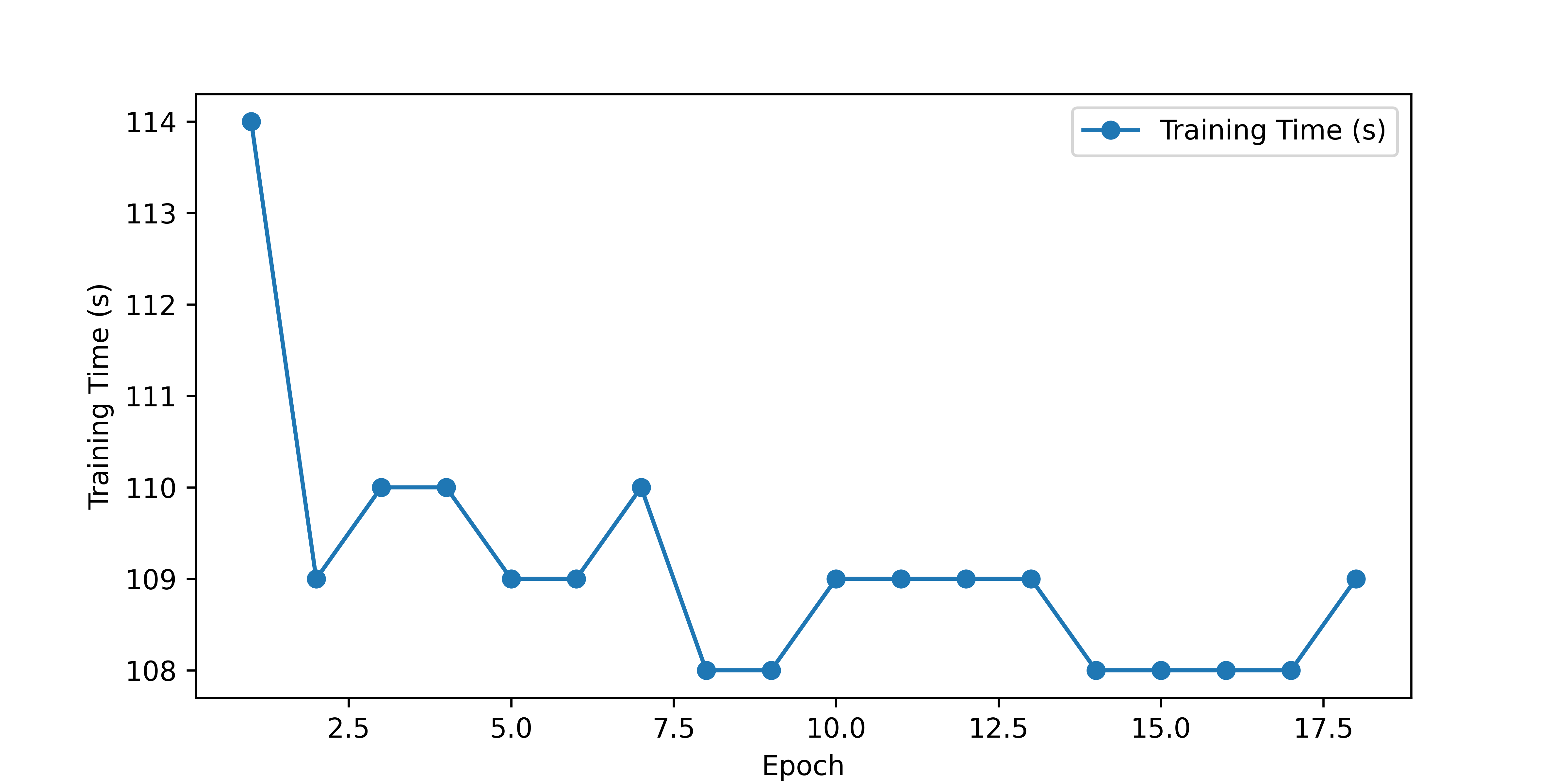}
        \caption{BloodMNIST}
        \label{subfig:sub1}
    \end{subfigure}
    \hfill
    \begin{subfigure}{0.42\textwidth}
        \includegraphics[width=\linewidth]{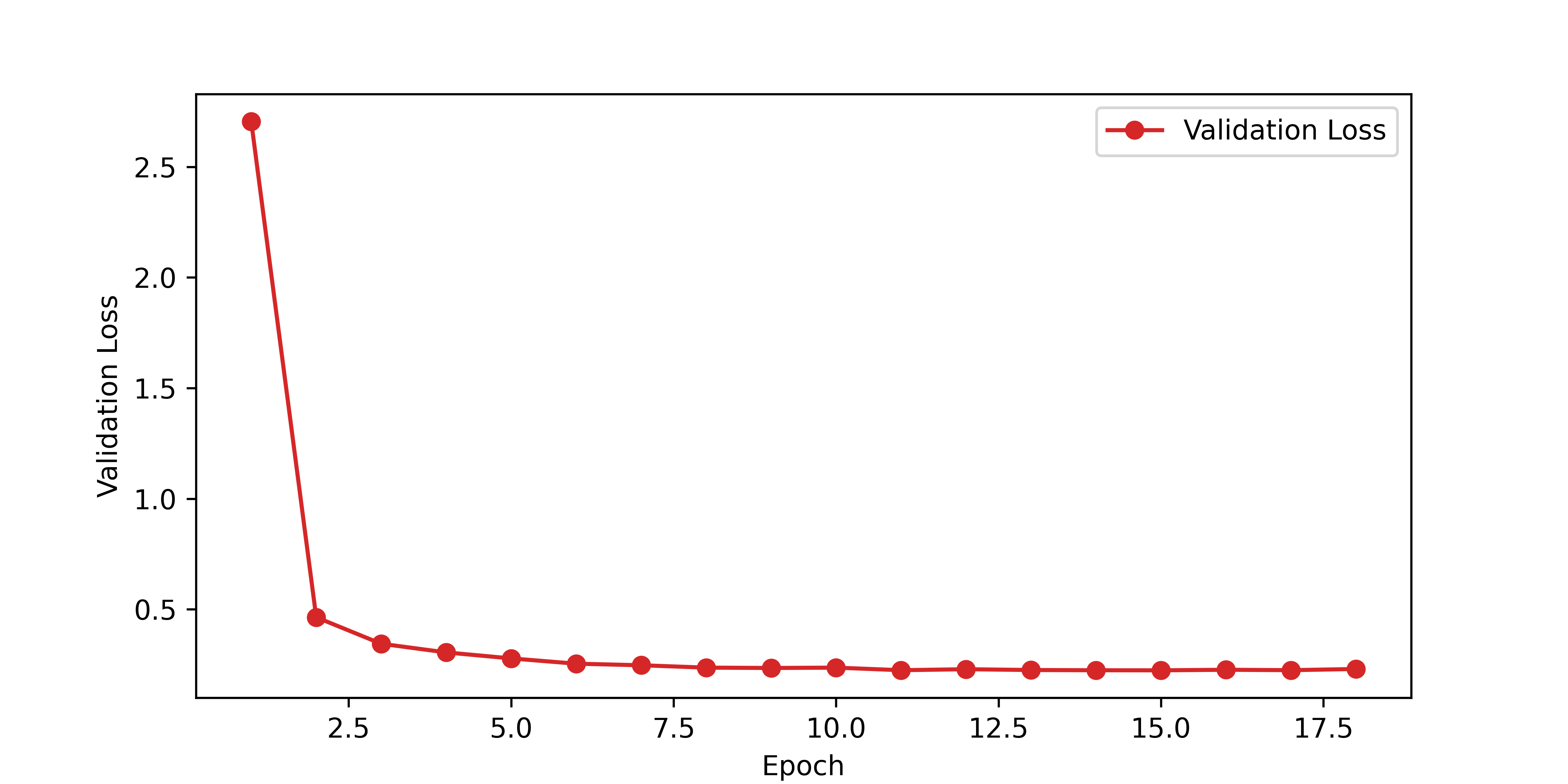}
        \caption{BloodMNIST}
        \label{subfig:sub2}
    \end{subfigure}
\hfill
\begin{subfigure}{0.42\textwidth}
        \includegraphics[width=\linewidth]{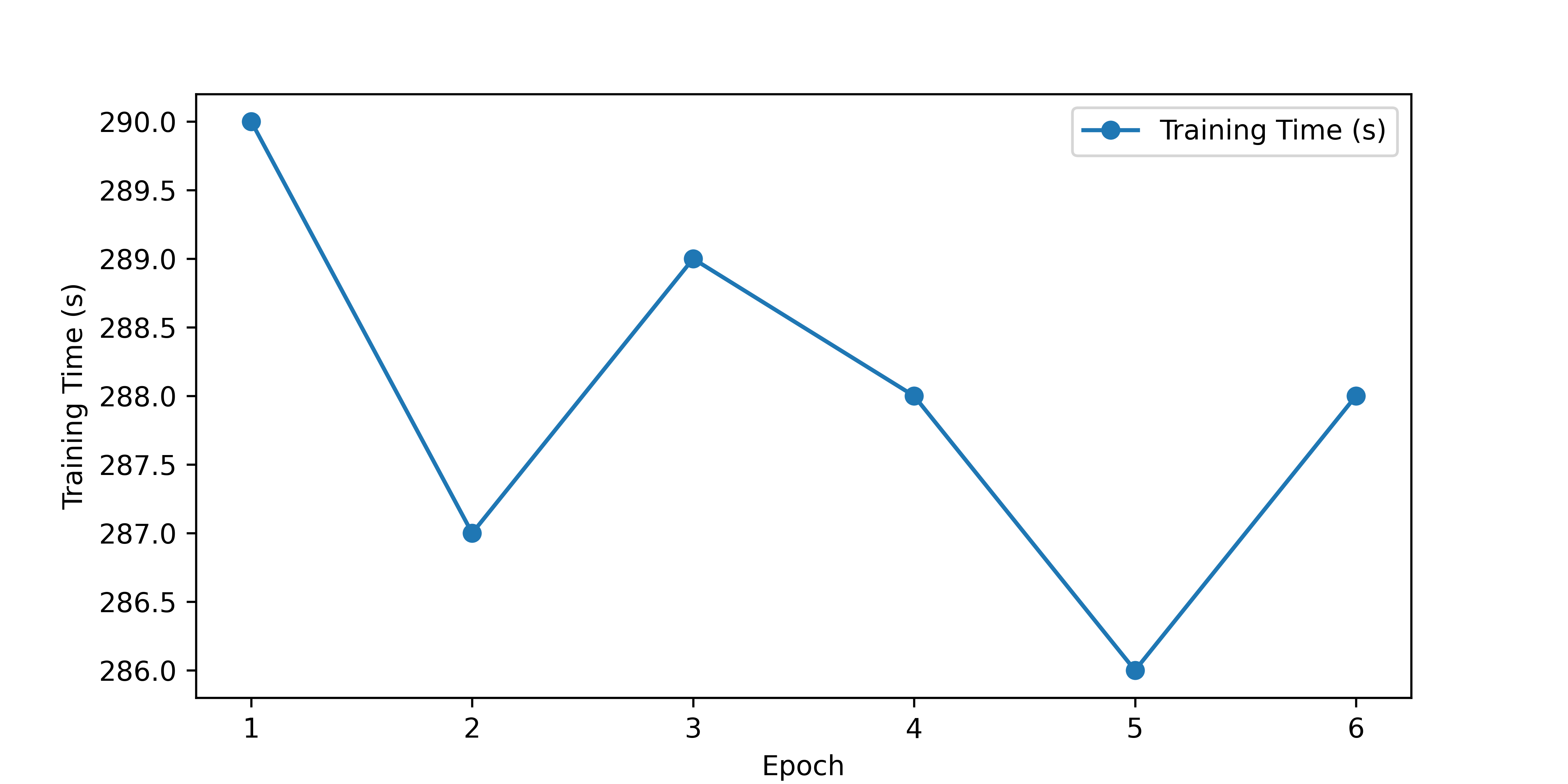}
        \caption{OrganAMNIST}
        \label{subfig:sub1}
    \end{subfigure}
    \hfill
    \begin{subfigure}{0.42\textwidth}
        \includegraphics[width=\linewidth]{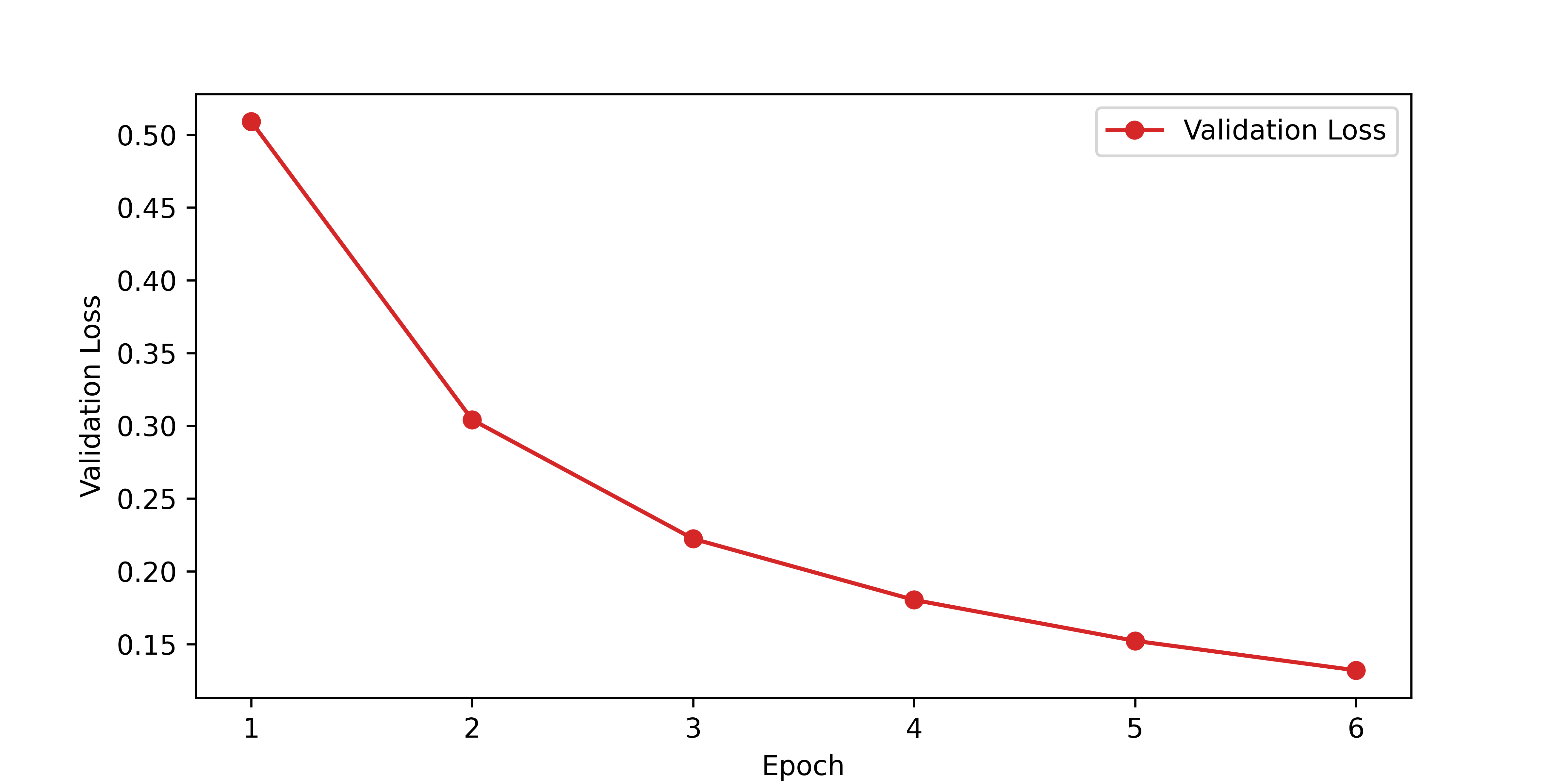}
        \caption{OrganAMNIST}
        \label{subfig:sub2}
    \end{subfigure}
\hfill
\begin{subfigure}{0.42\textwidth}
        \includegraphics[width=\linewidth]{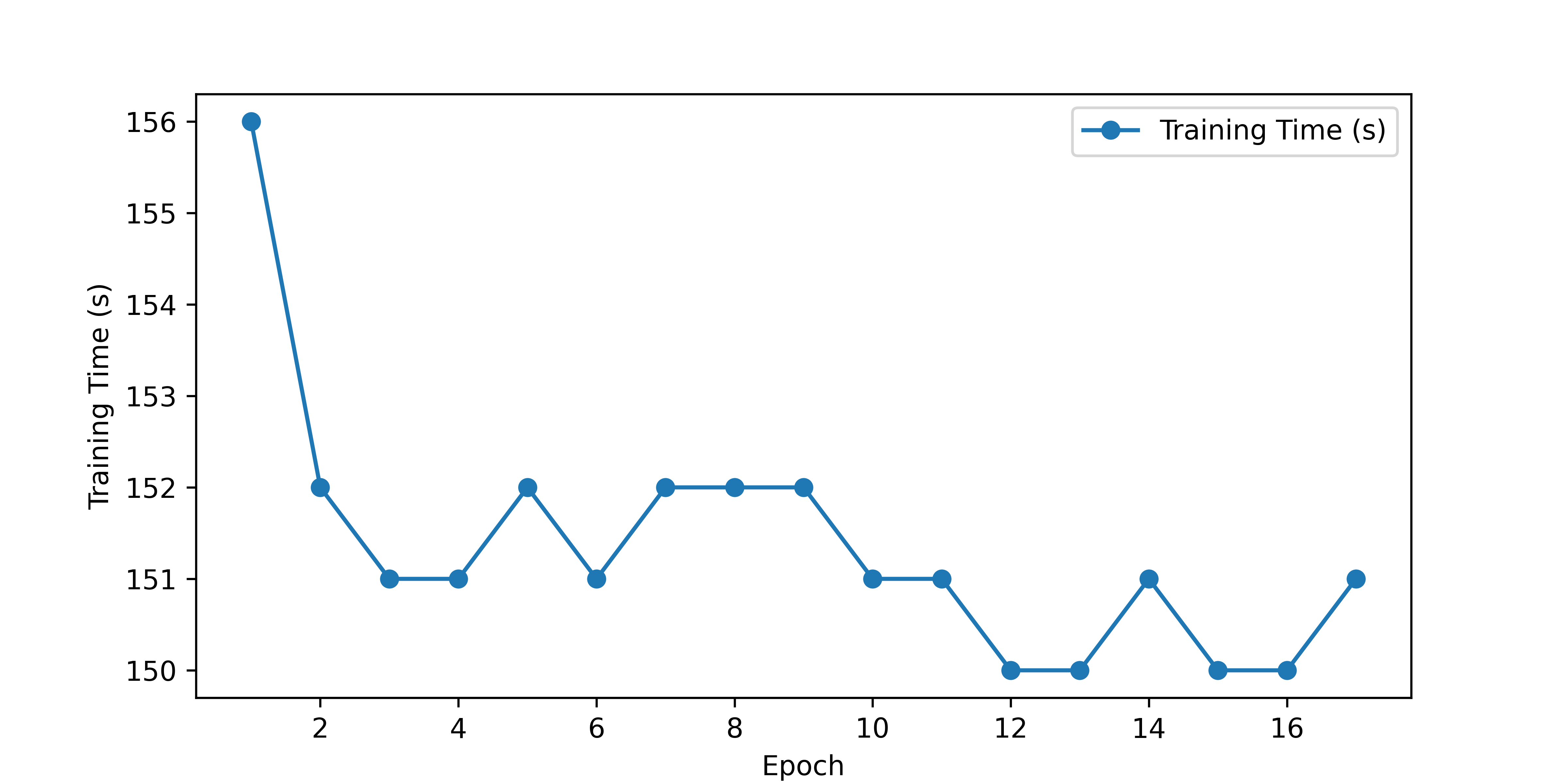}
        \caption{OrganCMNIST}
        \label{subfig:sub1}
    \end{subfigure}
    \hfill
    \begin{subfigure}{0.42\textwidth}
        \includegraphics[width=\linewidth]{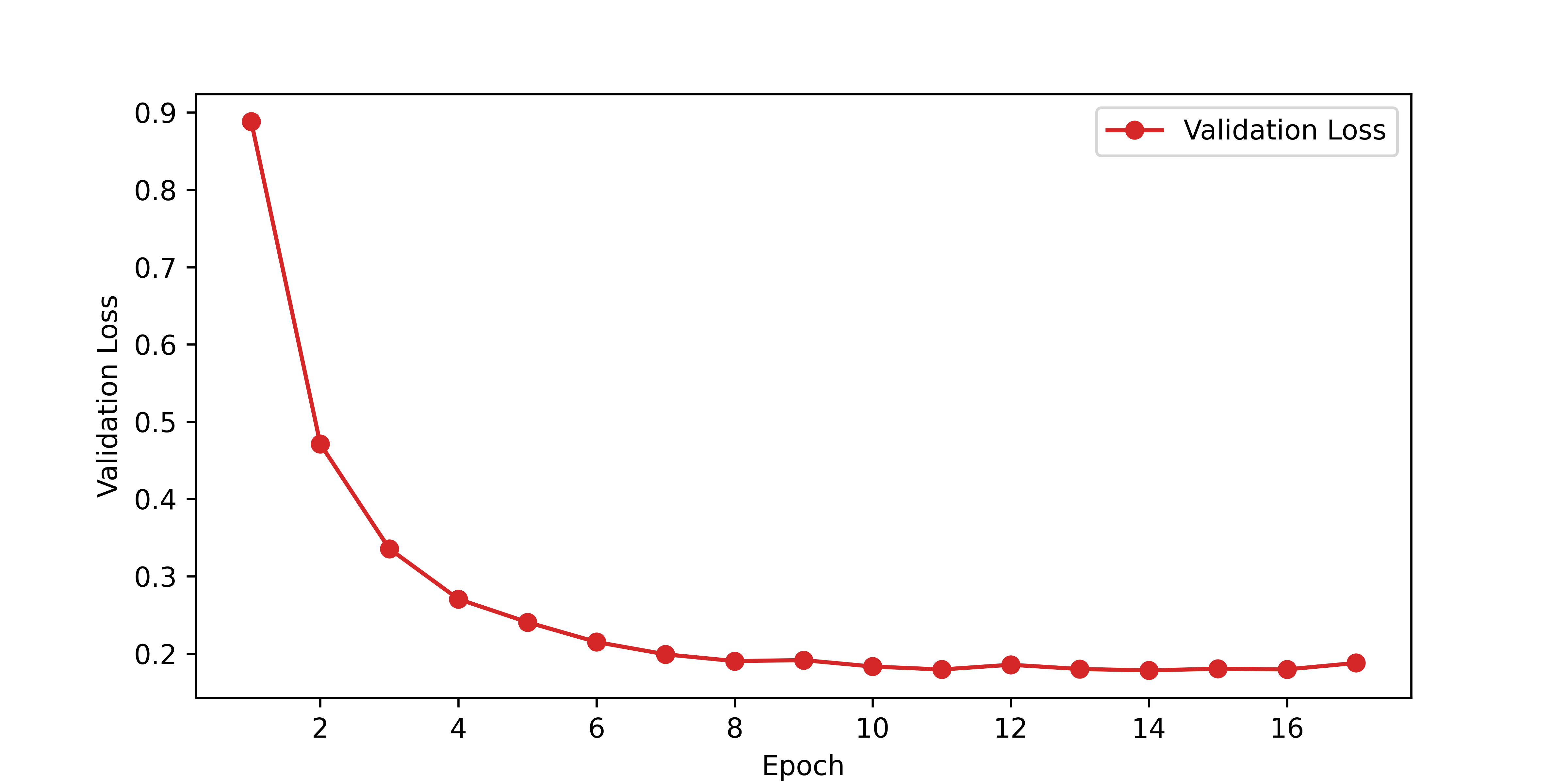}
        \caption{OrganCMNIST}
        \label{subfig:sub2}
    \end{subfigure}
\hfill
\begin{subfigure}{0.42\textwidth}
        \includegraphics[width=\linewidth]{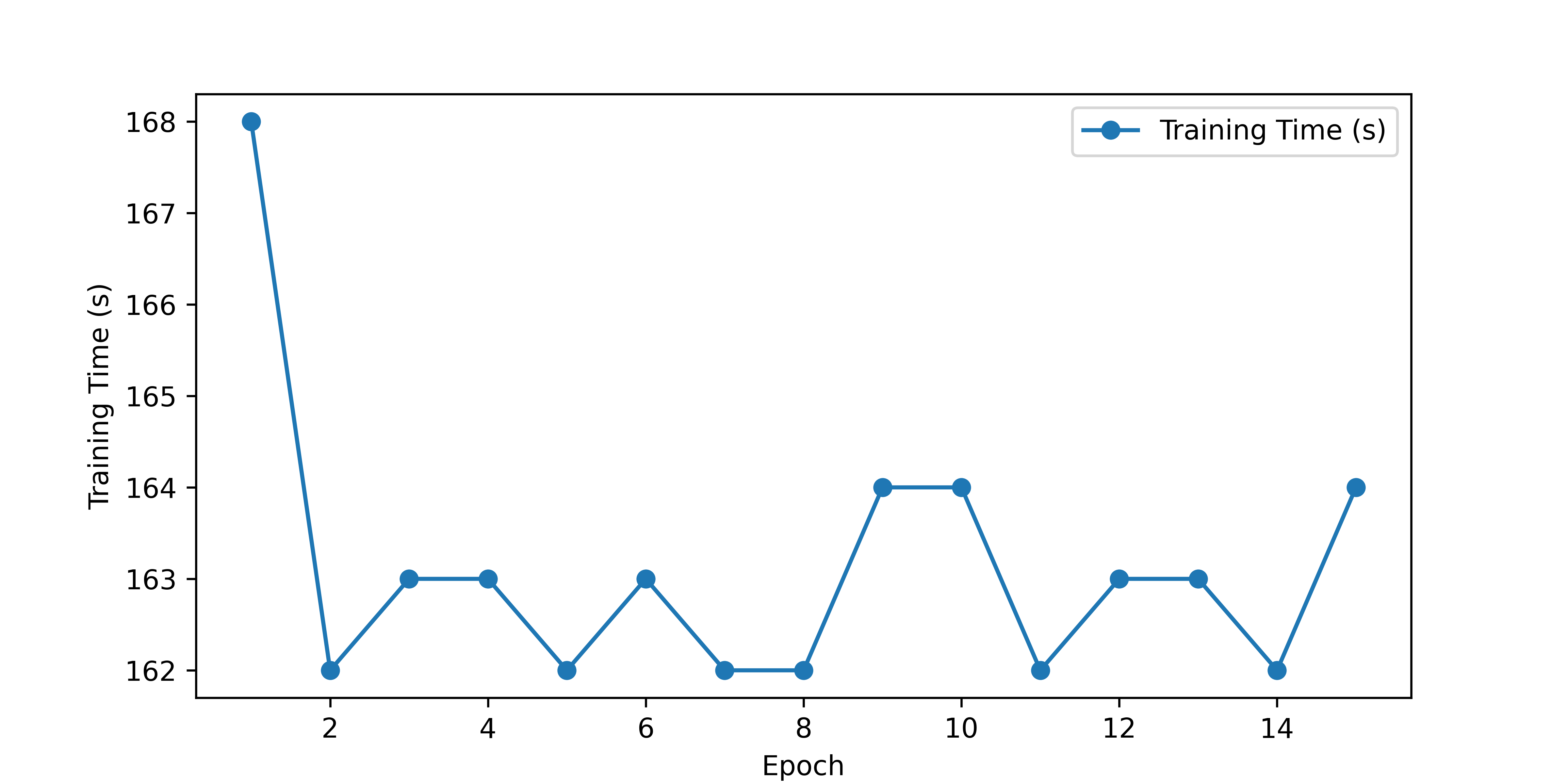}
        \caption{OrganSMNIST}
        \label{subfig:sub1}
    \end{subfigure}
    \hfill
    \begin{subfigure}{0.42\textwidth}
        \includegraphics[width=\linewidth]{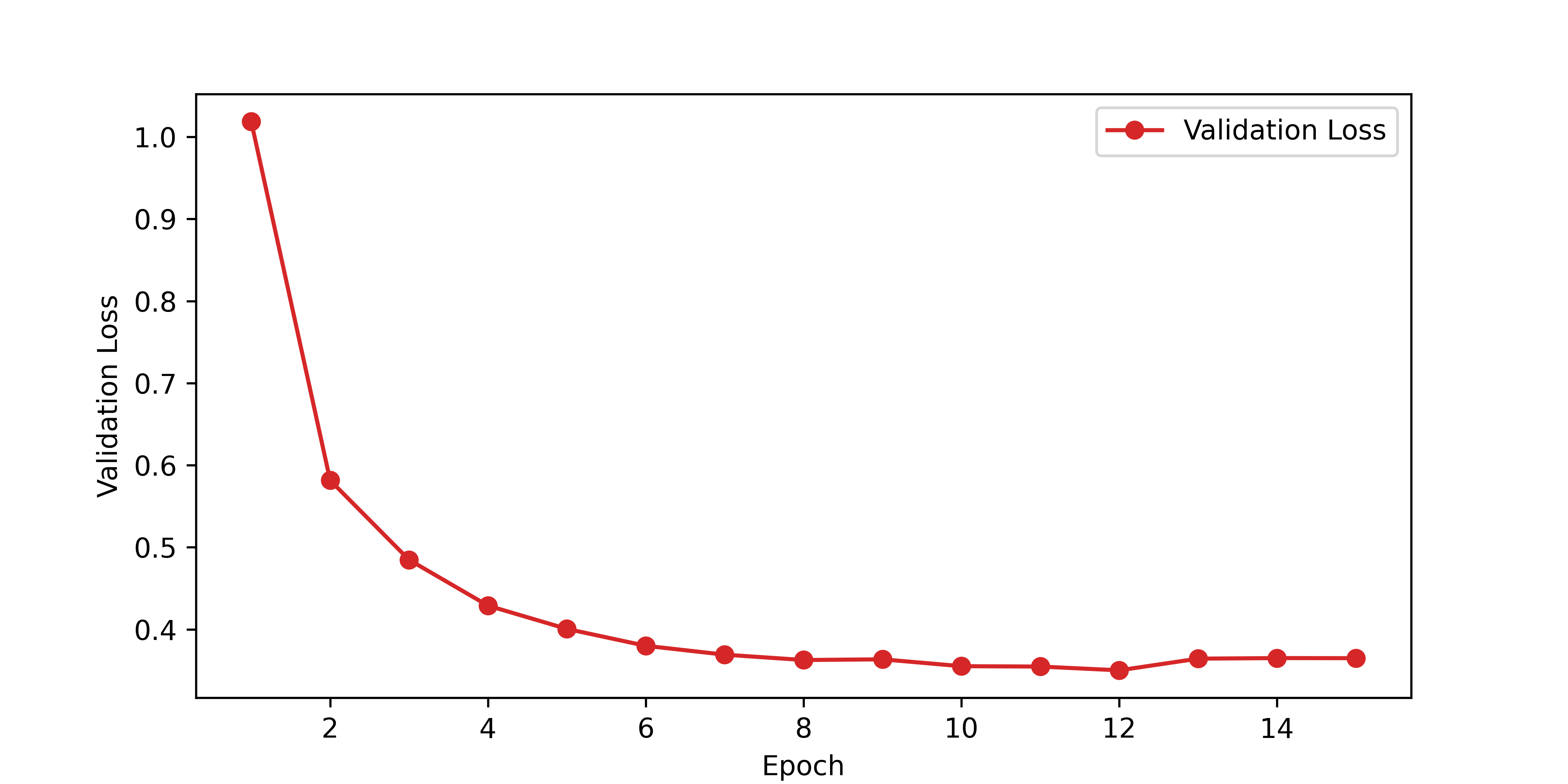}
        \caption{OrganSMNIST}
        \label{subfig:sub2}
    \end{subfigure}

    \caption{{\papername} performance on various datasets}
    \label{fig:main}
\end{figure*}

The results in Table \ref{tab:results} and Figure \ref{fig:main} demonstrate the performance of {\papername} in medical image classifications while protecting privacy. Encrypted models achieve test accuracy within 1-2\% of unencrypted baselines across datasets like DermaMNIST and BloodMNIST. For example, encryption incurs a negligible 0.15\% drop in accuracy on BloodMNIST relative to the 91.17\% unencrypted performance. Thus, the underlying model utility is preserved for training and inference after applying encryption. However, additional computation time is required for encrypted training, resulting in over 30$\times$ longer training times because of FHE computations. Training on BloodMNIST requires 33.56 minutes with encryption versus just 59.73 seconds without. So in terms of accuracy-privacy tradeoffs, the {\papername} narrowly limited accuracy reductions on sensitive patient data while upholding robust privacy guarantees.
\begin{table*}[!ht]
\centering
\caption{Performance of \papername}
\begin{tabular}{cccccc}
\toprule

 \textbf{Dataset} & \textbf{Enc training time} & \textbf{Enc Accuracy} & \textbf{Unenc accuracy} & \textbf{Unenc time}  & \textbf{\#Epochs} \\
\midrule
        DermaMNIST & 20.96 mins & 76.06\% & 76.16\% & 30.25 s & 12 \\
        BloodMNIST &        33.5611 mins           &       91.32\%           &             91.17\%            &      59.73 s   & 18\\
        OrganAMNIST &         30.2591 mins               &        88.59\%                 &               88.70\%       &       62.20 s      & 6 \\
        OrganCMNIST &        44.5961 mins                  &         88.20\%              &                88.26\%         &        77.94 s         & 17\\
        OrganSMNIST &           42.3798 mins                   &        75.94\%              &                76.26 \%         &        75.63 s     & 15\\
        
\bottomrule
\end{tabular}
\label{tab:results}
\end{table*}


\begin{table*}[!ht]
\caption{Training parameters}
\centering
\begin{tabular}{c c c c}
\toprule
\textbf{Dataset} & \textbf{\#Epochs} & \textbf{Learning rate} & \textbf{Batch size} \\
\midrule
        DermaMNIST & 12 & 0.01 & 512 \\        

        BloodMNIST  &         18        &         0.1              & 512  \\ 
        OrganAMNIST  &         6        &        0.01               &  512 \\
        OrganCMNIST  &         17        &        0.01               & 512 \\
        OrganSMNIST  &         15        &        0.01            &  512 \\
\hline
\end{tabular}
\label{tab:parameters}
\end{table*}
The training configurations used to benchmark performance are outlined in Table \ref{tab:parameters}. Hyperparameters like learning rate and batch size are tuned per dataset to optimize accuracy-privacy tradeoffs. 
In total, the experiments demonstrate encrypted medical imaging pipelines can deliver high test accuracy while protecting patient privacy, although further optimizations could continue improving runtime. Ethical and responsible development of such privacy-preserving machine learning techniques remains essential for realizing the benefits of AI in healthcare without compromising patients' privacy.

While encrypted computing currently entails a performance gap compared to unencrypted approaches, it excels in contexts where preserving data privacy is the paramount priority. The slower speed may prove a worthwhile tradeoff to guarantee privacy protections for sensitive user data. As encryption techniques continue to advance, performance costs will concomitantly lessen. For now, the privacy assurances encrypted computing affords already open up important, privacy-centric use cases that would otherwise remain infeasible.

\subsection{Comparison}
Table \ref{table:comparison} provides a thorough accuracy comparison between our proposed approach, {\papername}, and the state-of-the-art models introduced in recent research by \cite{medmnistv2} on medical image datasets. It is worthwhile to note that, their approach operates on plain data, while ours operates solely on encrypted data. This section serves to highlight the comparative standing of encrypted fine-tuning using {\papername} against training on plain data.

\begin{table*}[!htbp]
\caption{Comparison of {\papername} with state-of-the-art models}
\centering
\begin{tabular}{ccc}
\toprule
\textbf{Dataset} & \textbf{Accuracy of {\papername}} & \textbf{Accuracy of \cite{medmnistv2}}\\
\midrule
        DermaMNIST &       76.06\%              &  76.8\%\\        

        BloodMNIST   &       91.32\%         &       99.8\% \\
        OrganAMNIST  &      88.59\%            &        95.1\% \\
        OrganCMNIST  &      88.20\%          &        92\%\\
        OrganSMNIST   &      75.94\%          &      81.3\%  \\
\bottomrule
\end{tabular}
\label{table:comparison}
\end{table*}

For the DermaMNIST dataset, both {\papername} and the model proposed by \cite{medmnistv2} show similar accuracy results, indicating comparable performance. For BloodMNIST, although {\papername} exhibits slightly lower accuracy compared to \cite{medmnistv2}, it still achieves a commendable accuracy. Moreover, in the OrganAMNIST, OrganCMNIST, and OrganSMNIST datasets, {\papername} demonstrates competitive accuracy levels in comparison to \cite{medmnistv2}. This highlights the potential effectiveness of {\papername} in privacy-preserving ML for medical image classification.

\section{Conclusion}
In this paper, we present {\papername}, a fine-tuning framework designed for privacy-preserving ML on homomorphically encrypted medical image data. Our experiments demonstrate the strong performance of {\papername} in ensuring privacy while maintaining accuracy, with minimal deviation from unencrypted models. Additionally, comparative analysis with state-of-the-art models indicates that {\papername} has the potential to achieve state-of-the-art results in medical image classification. In our future work, we will focus on more complex medical image datasets.
\section*{Acknowledgement}
This work was supported in part by the National Science Foundation under Grants 2020636, 2054968, 2118083, 2315596, and 2244219 and in part by the Microsoft Faculty Fellowship Program.

\bibliography{aaai24}

\end{document}